# Deep learning for flash drought forecasting and interpretation

Qian Zhao[1]✉, Xuwei Tan[2], Xueru Zhang[2], Pierre Gentine[3,4], Yanlan Liu[1]✉

## Affiliations

[1] School of Earth Sciences, The Ohio State University, Columbus, OH, USA

[2] Department of Computer Science and Engineering, The Ohio State University, Columbus, OH, USA

[3] Department of Earth and Environmental Engineering, Columbia University, New York, NY, USA

[4] Center for Learning the Earth with Artificial Intelligence and Physics (LEAP), Columbia University, New York, NY, USA

✉Corresponding authors: Qian Zhao: zhao.4243@osu.edu, Yanlan Liu: liu.9367@osu.edu

## Abstract

Flash droughts are increasingly occurring worldwide due to climate change, causing widespread socioeconomic and agricultural losses. However, timely and accurate flash drought forecasting remains challenging for operational forecast systems due to uncertain representations of interactive hydroclimate processes. Here, we introduce the Interpretable Transformer for Drought (IT-Drought), a deep learning model based on publicly available hydroclimate data. IT-Drought skillfully forecasts soil moisture up to 42 days and flash droughts up to 11 days ahead on average across the contiguous U.S., substantially outperforming existing forecast systems limited to 10-day skillful soil moisture forecasting and unable to forecast flash droughts. IT-Drought offers novel data-driven interpretability on flash drought onset, benchmarking the memory effects of soil moisture, and revealing heterogeneous impacts and effective time ranges of climatic conditions—predominantly radiation and temperature. The findings highlight the potential of leveraging IT-Drought for early warning and mechanistic investigation of flash droughts, ultimately supporting drought risk mitigation.

**Main Text**

Droughts are among the most disastrous and costly natural hazards, causing impaired ecosystem functions and devastating socioeconomic losses[1-3]. Climate warming has accelerated drought onset and intensification, leading to a global transition to more flash droughts that unfold at subseasonal-to-seasonal (S2S) time scales, ranging from weeks to months[4,5]. Flash droughts pose a new challenge for drought risk management, as societal preparedness and mitigation strategies heavily rely on S2S forecasts to guide water resource planning, agricultural decisions, and early warning systems[6]. However, current process-based forecast systems struggle to reliably forecast flash droughts due to challenges in accurately representing the interactive and rapidly evolving hydroclimate dynamics that drive their onset[7,8]. For example, besides precipitation deficits, flash droughts can also be triggered by high evapotranspiration (ET) induced by heat that rapidly depletes soil moisture (SM), accelerating surface heating and drying through a positive land-atmosphere feedback[9,10]. The accuracy of process-based flash drought forecast is further limited by errors in initializing and representing the temporal memory (lag effect) of SM, which is key to drought forecasting but lacks robust data-driven benchmarks[11-15].

Artificial Intelligence (AI) models hold the potential to improve drought forecasts, benefiting from flexible representations of feature interactions and memory effects[16]. Recognizing this potential, previous studies have successfully applied deep learning models to forecast droughts by learning the hidden structures in the temporal evolution of drought indices[17-19]. More recently, advanced deep learning architectures, such as the Transformer algorithm[20], have demonstrated superior predictive performance, leveraging the attention mechanism that more effectively captures temporal dependencies across multiple time scales and predictive features[21,22]. However, beyond predictive accuracy, many existing AI-based drought forecasts lack interpretability, which obscures the understanding of the impacts of climate drivers and memory effects on drought onset[16]. Using empirical drought indices (e.g., standardized precipitation index) instead of physical variables (e.g., SM) also impedes bridging AI-exploited relations from data to process-level interpretation of drought development. Addressing these gaps through interpretable AI is essential for advancing the understanding and robust forecasts of flash droughts[3,6,23].

Here, we introduce an Interpretable Transformer for Drought (IT-Drought) to forecast flash droughts across the contiguous US (CONUS), while providing data-driven interpretability. IT-Drought is based on an encoder-decoder Transformer architecture (Fig. 1) to forecast the target sequence of SM at a 46-day forecast horizon, using historical sequences of daily meteorological features and SM over the past year as inputs. IT-Drought views drought forecasting as a regression task that forecasts SM, allowing flexibility in the thresholds used to identify flash droughts while facilitating the interpretation of hydroclimatic factors leading to flash droughts. We then detect flash droughts as episodes when SM falls below its 40$^{th}$ percentile and then recovers above its 20[th] percentile, with durations from one to three months[4]. IT-Drought further employs Integrated Gradients (IGs)[24], a gradient-based attribution method, to quantify the contribution of antecedent SM and climate features to the variation in SM during the onset phase of flash droughts. We leverage the decay of IGs over time lags to understand the time scales of memory and climate impacts on flash drought onset.

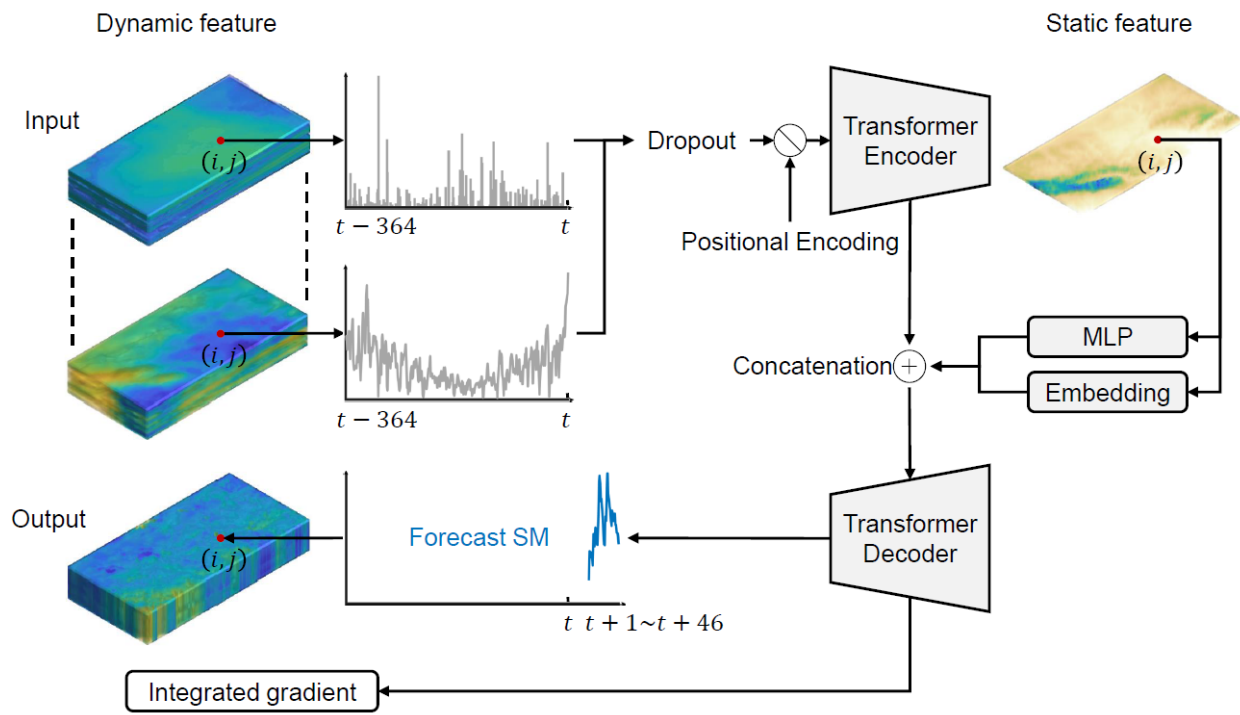

**Fig. 1 | The architecture of the Transformer-based interpretable deep learning model (IT-Drought).** IT-Drought forecasts soil moisture (SM) sequence by integrating dynamic and static features based on an encoder-decoder structure of a Transformer. IT-Drought achieves temporal learning using forecast windows, which learns from daily dynamic features over the past year and generates SM sequences for the upcoming 46 days simultaneously in each forecast window. IT-Drought interprets the impact of antecedent dynamic features on forecasted SM by applying the Integrated Gradient to the trained model.

**Predictability of soil moisture and flash droughts**

IT-Drought is trained, validated, and tested using historical reanalysis records from 1979 to 2022, with the testing period used to report model accuracy (Methods). The accuracies of SM and flash drought forecasts are measured using the anomaly correlation coefficient (ACC) and F1 score, respectively. We compare the performance of IT-Drought with that of five existing process-based S2S forecast systems, including the Global Ensemble Forecast System (GEFS) and the S2S prediction project hosted by the European Centre for Medium-Range Weather Forecasts (ECMWF, Extended Data Table 1).

IT-Drought forecasts SM and flash droughts more accurately than existing S2S forecast systems. At lead times ranging from 1 to 46 days, IT-Drought forecasts daily SM with ACC varying from 0.98 to 0.48 on average across CONUS, outperforming existing forecast systems showing an average ACC from 0.58 to 0.14 (Fig. 2a). Despite decreasing forecast accuracy at longer lead times, IT-Drought still shows skillful SM forecasts (ACC > 0.50) at lead times up to 42 days, substantially extending the 10-day skillful forecast range of the best-performing GEFS. While existing forecast systems struggle to forecast flash droughts with F1 scores below 0.32 at all lead times, IT-Drought provides skillful flash drought forecasts (F1 > 0.50) at lead times up to 11 days, averaged across CONUS (Fig. 2b). IT-Drought balances between precision and recall in flash drought forecasts, varying from 0.83 to 0.55 and from 0.84 to 0.54, respectively, at lead times from 1 to 10 days (Extended Data Fig. 1). The improvement of IT-Drought over forecast systems becomes limited beyond three weeks. The F1 score of IT-Drought drops below 0.30 at lead times exceeding 3 weeks, comparable to 0.14-0.21 of forecast systems. The limited predictability of flash drought at long lead times highlights the challenge of forecasting a chaotic climate.

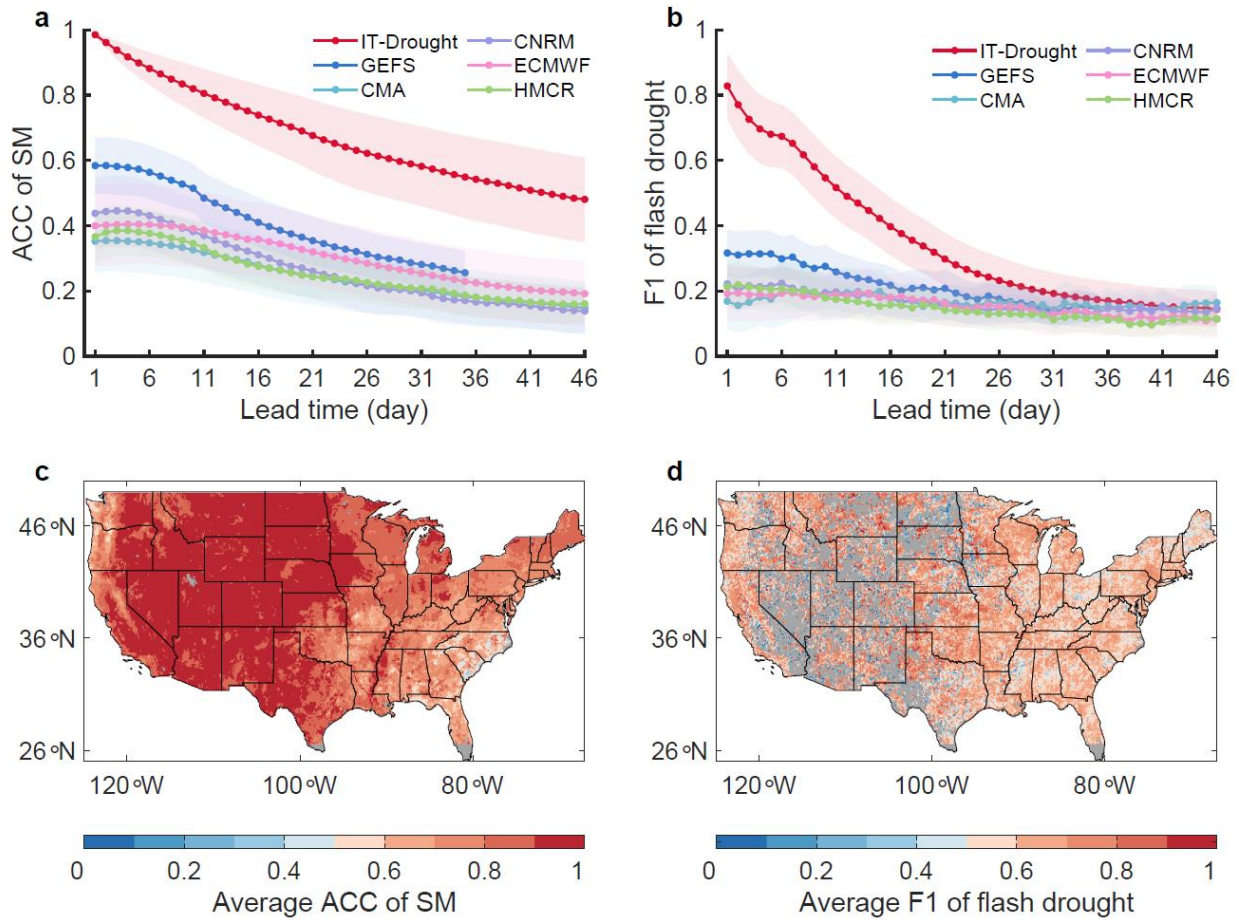

**Fig. 2 | IT-Drought outperforms existing dynamic forecast systems in forecasting soil moisture (SM) and flash droughts.** The anomaly correlation coefficient (ACC) of root-zone SM (**a**) and the F1 score of flash drought (**b**) forecasted by IT-Drought and five dynamic forecast systems (GEFS, ECMWF, CMA, HMCR, and CNRM) at lead times from 1 to 46 days. The shaded areas show half of the standard deviations of the accuracies across CONUS. The spatial patterns of ACC (**c**) and F1 score (**d**), averaged over lead times of 1–14 days. Both ACC and F1 are evaluated for the testing period (Extended Data Table 1).

IT-Drought shows skillful SM forecasts in 99.56% of CONUS, where the average ACC over lead times up to two weeks exceeds 0.5 (Fig. 2c). Flash drought forecast is more challenging—84.25% of CONUS show two-week averaged F1 score exceeding 0.5 (Fig. 2d). This is because flash droughts frequently occur in humid regions, including the Eastern U.S. and the coastal Pacific Northwest (Extended Data Fig.2), where SM is less accurately forecasted than in dryer regions. The lower predictability of SM is attributable to stronger high-frequency variations of SM (Extended Data Fig. 3). Using a Fourier transform to analyze the spectrum, we find that ACC is negatively correlated with the spectral exponent of SM time series in each grid cell ($R^2 = 0.69$, $p < 0.001$), where a more negative exponent indicates stronger signals of low-frequency components in SM variability[25,26]. The result suggests that, compared to regions where SM is dominated by slowly varying components (e.g., seasonal cycles), SM forecasting is more challenging in regions where SM is driven by high-frequency signals, resulting from rapidly varying hydrological processes of infiltration, percolation, runoff, evaporation, and plant water uptake.

**Interpreting soil moisture memory**

The self-dependency on antecedent SM, i.e., SM memory, shows the largest positive contribution—one order of magnitude larger than those of climate variables—to the forecasted SM during flash drought onset (Fig. 3a). Surface SM also substantially contributes to the forecasted root-zone SM due to infiltration. The dominant contribution of SM memory to SM-based drought forecast is consistent with previous model-derived findings[11]. To quantify the time range of SM memory, we define $t_{SMM}$ as the number of antecedent days when the lagged SM substantially (IG > 0.002) affects the forecasted SM in flash droughts. On average across

CONUS, $t_{SMM}$ is estimated to be 16.36 days, which nonetheless shows large spatial variability (Fig. 3b). The Northeast, Midwest, and Northwest typically exhibit longer SM memory ($t_{SMM}$ > half a month) compared to the Southeast and the Southwest during flash drought onsets. Given that $t_{SMM}$ positively correlates with long-term average SM (Fig. 3c), the longer memory likely arises from a larger soil water storage that increases SM residence time based on soil water balance[27].

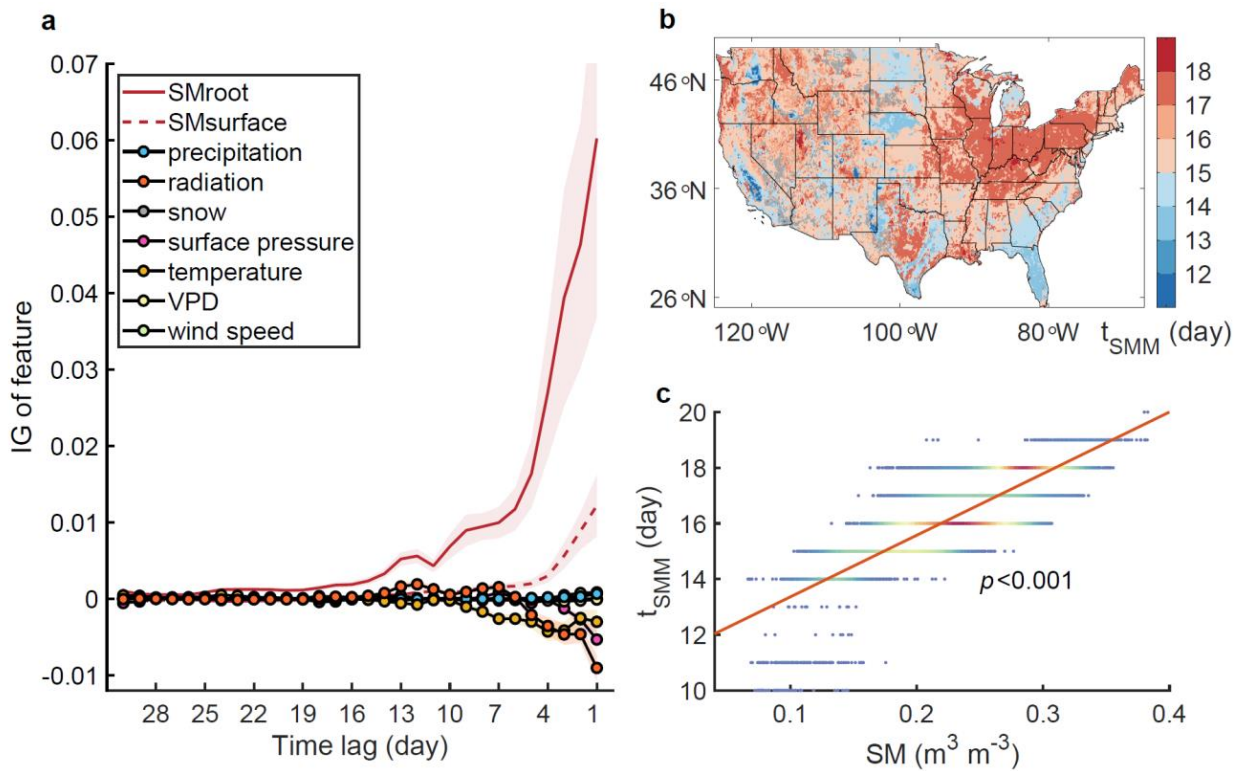

**Fig. 3 | Soil moisture (SM) memory provides the dominant source of predictability for flash drought. a**, The impacts of lagged surface SM (SMsurface) and root-zone SM (SMroot) and climate features on root-zone SM during flash drought onset, as measured by the Integrated Gradient (IG). The lines and shaded areas show the means and standard deviations across CONUS. **b**, The spatial pattern of the length of SM memory ($t_{SMM}$) derived from the IG decay curve of root-zone SM. **c**, The positive relationship between $t_{SMM}$ and the long-term average

root-zone SM across CONUS. Each dot represents a grid cell, with warmer colors representing higher point densities.

**Interpreting lagged climate impacts on flash droughts**

SM in flash droughts is regulated by antecedent climates, which show various impact strengths and effective time ranges (Methods). Overall, high shortwave radiation, high air temperature, and low precipitation contribute to reducing SM and thus forthcoming flash droughts (Fig. 4). Based on the IG magnitudes, antecedent shortwave radiation in the previous week shows the most substantial impact on SM among all climate variables. High antecedent radiation reduces SM through elevated ET, exhibiting larger impact in summer and spring than in other seasons (Fig. 4a). Antecedent air temperature most persistently reduces SM, with an average effective time range of 9 days and longer in summer and autumn (Fig. 4b). This persistent effect is partly attributable to the slower temporal variability of temperature compared to radiation (Extended Data Fig. 4). The impacts of radiation and temperature on flash drought onset are especially prominent in the Central U.S. due to regionally strong land-atmosphere interactions[28,29] (Fig. 4d, e). Although the effect of antecedent precipitation is largely transferred to SM memory, it still makes a non-negligible contribution to SM through hydrological routing and snowmelt processes that are not captured by SM memory. The corresponding impact is more apparent in winter and spring seasons, and in topographically complex and snow-affected regions (Fig. 4c, f).

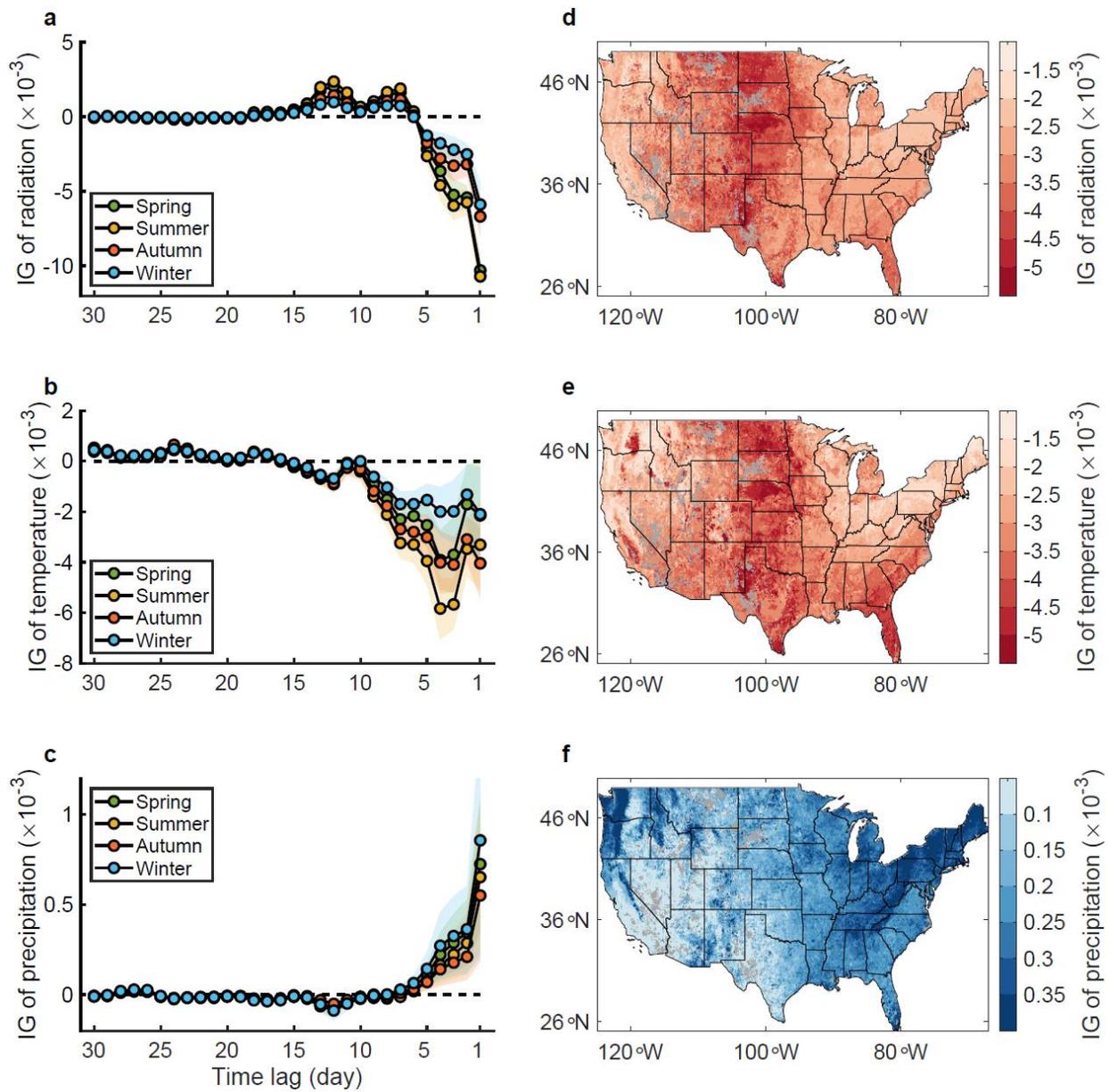

**Fig. 4 | Radiation, temperature, and precipitation exert substantial and heterogeneous impacts on soil moisture during flash drought onset.** The Integrated Gradients (IGs) of radiation (**a**), temperature (**b**), and precipitation (**c**) over time lags in the four seasons, respectively, showing decaying contributions over longer lead times. The dotted lines and shaded areas denote the means and standard deviations across CONUS. The spatial patterns of IGs of radiation (**d**), temperature (**e**), and precipitation (**f**) averaged over 7 days, showing regionally

strong impacts of radiation and temperature in the Central U.S., and of precipitation in snow-affected and topographically complex regions.

High surface pressure, high VPD, low wind speed, and low snowfall further contribute to depleting SM during flash drought onset (Extended Data Fig. 5). High antecedent surface pressure suppresses cloud formation, thereby exacerbating surface drying. The effect of surface pressure on SM is substantial—comparable to those of radiation and temperature—but short-lived (3 days, Fig. 3a). Antecedent VPD only shows minimal impact in humid regions, in contrast to its larger impact in the southwestern drylands. SM is further positively affected by wind speed through atmospheric moisture transport, particularly in the Central U.S., and by snowfall in snow-affected regions. These results reveal the heterogeneous magnitudes and time scales of climatic variables driving the onset of flash droughts.

**Discussion**

Our AI-based model substantially improves the forecasts of SM and flash droughts up to 46 days in advance compared to the state-of-the-art process-based forecast systems. Note that the accuracies of IT-Drought and forecast systems are subject to uncertainties in the SM product used as the benchmark, despite NLDAS-2 being found most representative in capturing droughts compared to other reanalysis products[30]. The relative performance of the examined forecast systems also varies with the benchmarking product. GEFS shows the highest skill when benchmarked with NLDAS-2 SM due to their shared model components (Fig. 2a). When benchmarked with ERA5-land SM instead, the performance of ECMWF and CNRM improves given their shared model components and source of SM initialization (Extended Data Fig. 6).

Nevertheless, the large improvement of IT-Drought over best-performing reanalysis forecast systems (Fig. 2) highlights the limitation of process-based S2S forecasts due to uncertainties in forcing, initialization, parameterizations, and structural representations of climate, surface hydrology, plant physiology, and land-atmosphere interactions[16,31]. These uncertainties amplify errors in estimated SM, particularly the high-frequency components[32], thereby challenging the forecast of rapidly varying SM during flash droughts. Using an alternative approach without preconceived assumptions on hydrological and land-atmosphere interactions, IT-Drought emulates these processes by exploiting the interactions between climate and SM data and their dynamic lagged effects. IT-Drought effectively exploits such interactions from long time series encompassing ample flash drought events, which enables skillful forecasts of SM and flash droughts.

Although IT-Drought improves SM and flash drought forecasts, accurate forecasts remain challenging in regions exhibiting strong high-frequency variability and short SM memory, especially at lead times beyond a month (Figs. 2, 3, Extended Data Fig. 3). The challenge stems from the fast dissipation of predictive signal in lagged antecedent conditions. Signal dissipation is likely exacerbated by precipitation showing large high-frequency variability, which weakens its S2S predictability and thus SM and flash droughts forecasts[32-34]. Additionally, rapid variability of SM can also arise from dynamics at faster time scales, such as diurnal plant water use and atmospheric feedback, and processes relatively independent from climate, such as irrigation[35-37]. Future improvement of our flash drought forecast system will likely benefit from ensemble forecasts[38], improved data representation of human activities, and hybrid physics-AI modeling that better constrains physical and biological dynamics[39-41].

Beyond achieving high forecasting accuracy, a key to leveraging AI for advancing the understanding of hydroclimate extremes is model interpretability[16,42-44]. IT-Drought allows extracting interpretable information based on IG, including SM memory and lagged climate effects, providing a data-driven benchmark for understanding flash drought onset. The dominant role of SM memory highlights the importance of correctly characterizing antecedent SM up to a month for flash drought forecasts. The length and spatial patterns of SM memory estimated here (Fig. 3) are comparable to previous estimates derived from remote sensing and process-based models, suggesting shorter SM memory in the Western U.S. during flash drought onset and water-limited soil dry-down phases[45,46]. The short SM memory identified here is dominated by small soil water storage rather than fast loss through ET, as locations exhibiting short SM memory tend to show slower water loss during flash drought onset due to water-limited ET (Extended Data Fig. 7). Our AI-derived memory, $t_{SMM}$, shares a similar concept with previously used autocorrelation-based memory metrics[15,25,47,48], while being free from the assumption that SM dry-down follows a first-order Markovian process. It allows for estimating the self-dependency of SM while considering time-varying climate impacts, providing a more coherent data-driven benchmark that better informs the impact of antecedent SM.

IT-Drought reveals the impacts of climate conditions on flash drought, which are consistent with physical understanding, while providing deeper insights into their relative contributions, spatial heterogeneity, seasonal variability, and effective time ranges. Because flash droughts occur more frequently in humid regions (Extended Data Fig. 2), where ET before flash droughts is typically energy-limited[1], antecedent radiation dominates SM depletion during flash drought onset. High temperature elevates ET through increased evaporative demand and stomatal conductance due to enhanced photosynthesis before a thermal threshold, and increased aerodynamic conductance due to enhanced boundary layer mixing[49,50]. Although high temperature could also promote ET via increased VPD[51], IT-Drought suggests that antecedent VPD only marginally influences flash drought onset. This is because ET before flash droughts is primarily driven by radiation rather than aerodynamics, particularly in humid regions, consistent with recent mechanistic findings[52]. The impact of VPD on SM decline is further attenuated by stomatal closure[53,54]. In topographically complex and snow-affected regions, the strong impact of antecedent precipitation and snow supports regional case studies suggesting snowmelt and lateral flow are critical processes regulating flash droughts during winter and spring seasons[55,56]. The findings highlight that considering the heterogeneous and coupled impacts of hydrological dynamics, land-atmosphere interactions, and physiological responses is essential for improving the forecast and mechanistic understanding of flash droughts using process-based models[57,58].

In summary, our AI-based model, IT-Drought, substantially improves the forecast accuracy of SM and flash droughts across the continental U.S. compared to existing process-based forecast systems. IT-Drought effectively exploits SM memory and lagged climate impacts in line with process-based principles, indicating that the high forecasting accuracy of IT-Drought stems from physically valid predictive sources. Benefiting from its capability to unravel complex time-varying interactions among predictive sources, IT-Drought provides a detailed data-driven benchmark on the dependency of flash drought onset on antecedent SM and climate forcings, which will support further research on flash drought mechanisms. The findings and tools established here set the foundation for operational early warning of flash droughts for enhanced drought risk management.

## Methods

**Input and target data**

The target of IT-Drought, root-zone soil moisture (SM, 0-100 cm), was obtained from the North American Land Data Assimilation System 2 (NLDAS-2) with 3-hour and 0.125° resolutions. The inputs consist of eleven dynamic features and seven static features. Dynamic features include climate variables of precipitation, 2-m temperature, surface pressure, vapor pressure deficit (VPD), wind speed, snowfall, and shortwave incoming radiation. Precipitation data were obtained from the Multi-Source Weighted-Ensemble Precipitation (MSWEP) version 2.80 with daily and 0.1° resolutions, which provides a high-quality product by merging gauge, satellite, and reanalysis data[1,2]. All other climate variables were obtained from the ERA5-land reanalysis dataset with hourly and 0.1° resolutions. Besides climatic variables, dynamic features also include the antecedent surface (0-40 cm) and root-zone SM to represent the memory effects. All the dynamic features are readily available and updated near real-time, allowing the use of IT-Drought for timely SM forecasts at S2S timescales. We further considered static features characterizing the physical and ecological conditions of the land surface, which could modulate SM dynamics. The static features include one categorical variable, land cover type from the National Land Cover Database[3], and six numerical static features, including topography from the Shuttle Radar Topography Mission[4], canopy height from the Global Forest Canopy Height 2019[5], and the mean and standard deviation of surface and root-zone SM for 1979-2022 from NLDAS-2. All dynamic features were aggregated into daily and 0.125° resolutions, and static features were aggregated into 0.125° resolutions. All inputs and the target were normalized to a range between 0 and 1 using min-max normalization.

**IT-Drought model**

The architecture of IT-Drought is illustrated in Fig. 1. IT-Drought uses a standard Transformer[6] with an encoder-decoder structure to forecast the target (root-zone SM) sequence. IT-Drought achieves temporal learning by exploiting the relationship between the historical sequence of input features and the target in a forecast window, using a Transformer encoder with a linear transformation and positional encoding. For each forecast window, IT-Drought uses a 365-day historical sequence of dynamic features (detailed below) to forecast the target sequence in the upcoming 46-day forecast window, a time range consistent with S2S forecasts of the ECMWF Integrated Forecasting System. To account for the spatially heterogeneous response of SM to inputs, IT-Drought learns spatial representations from static features using Multi-Layer Perceptrons and embedding techniques. The temporal and spatial representations are concatenated at each timestamp, which is fed into the Transformer decoder to generate the target sequences. The model architecture extends our previous work[7], which has demonstrated improved performance of Transformer-based forecasts compared to other commonly used deep-learning models. Here, we adapted the model to target SM prediction with near-real-time physical variables to provide daily forecasting of SM and flash droughts and leveraged interpretability for insights into flash drought development as described below.

**Training and experiments**

IT-Drought was trained on 32 years (1979-1994 and 2001-2016), validated on 4 years (1995-1996 and 2017-2018), and tested on 8 years (1997-2000 and 2019-2022) of data, respectively. We left a 2-year gap between the training and testing periods to remove the possible impacts of SM memory in the training periods on the testing periods[8]. During the training periods, we employed a 0.1 dropout rate on the training samples to prevent overfitting. The model was trained on 2900 batches with a batch size of 32, using a loss function defined as the mean absolute error of SM. Based on the trained IT-Drought, we evaluated its accuracy for SM sequences and flash drought events during the testing periods. We then used the trained IT-Drought to interpret the SM variation during the onset phase of flash droughts by quantifying its temporal dependency on each input feature as described below.

**IT-Drought interpretation**

To quantify the temporal dependencies of the target on input features, IT-Drought employs a gradient-based post-hoc attribution method, the Integrated Gradient (IG)[9]. IG has been successfully applied in previous studies to interpret hydrologic dynamics exploited by deep networks[10-12]. Formally, let $F$ represent the deep network that predicts SM, $x^t$ be the input at time $t$ that consists of eleven dynamic features over the 365 days before time $t$ and seven static features, and $x'$ be the baseline input that is typically a vector of zeros. IG computes the path integral of the gradients along the linear path from the baseline $x'$ to the input $x^t$.

$$IG(x_i^t) = (x_i^t - x_i') \times \int_0^1 \frac{\partial F(x' + \alpha \times (x^t - x'))}{\partial x_i^t} d\alpha \qquad (1)$$

where $x_i'$ and $x_i^t$ are the $i$th component in the $x'$ and $x^t$, respectively, and $\frac{\partial F(x^t)}{\partial x_i}$ is the gradient of $F(x^t)$ along the $i$th dimension. Here, a zero baseline was used for all dynamic features[12].

For each day during the onset phase of flash droughts, we calculated the IG with respect to each feature on each of the past 365 days to get a feature-specific curve of IG over lagged days. The IG curves for all drought onset days were then averaged to characterize the lagged impacts of input features on SM in each grid cell. For each feature, we quantified the magnitude and time scale of the lagged impact using the average IG over lags up to a week and the effective time range, respectively. The effective time range was defined as the number of antecedent days when the lagged feature substantially affects the forecasted SM, with IG > 0.002 for root-zone SM and absolute IG > 0.001 for other features. The effective time range of the root-zone SM represents the length of the SM memory.

**Process-based forecast systems**

IT-Drought was compared with baselines of operational process-based forecast systems in terms of their accuracies in forecasting SM and flash droughts. We analyzed the baseline products of daily averaged forecasts of the top 100 cm SM from GEFS, CMA, CNRM, ECMWF, and HMCR (details shown in Extended Data Table 1). GEFS was obtained from ref[13], and CMA, CNRM, ECMWF, and HMCR were obtained from the S2S database archived in ECMWF (https://apps.ecmwf.int/datasets/data/s2s/). We used reforecasts rather than historical forecasts of SM as baselines because of their longer data records and consistent model versions throughout the forecast periods. We refer to the reforecast skills of the forecast systems as forecast skills in comparison with those of IT-Drought. The forecast systems provide daily SM reforecasts at weekly or semi-weekly update frequencies with maximum lead times ranging from 35 to 61 days.

**Flash drought identification**

To assess the forecast skill for flash droughts, we identified flash droughts based on SM from NLDAS-2, IT-Drought, and baseline forecast systems. Similar to previous studies[14,15], we detected flash droughts based on weekly averages to reduce noise. To minimize the impact of systematic biases in SM across models on assessing drought forecast skills[16], we detected flash droughts after converting weekly SM to percentiles in each grid cell. The SM percentile of each week in a given year was determined as the percentile among all weekly SM in the same week across all years, i.e., 1979-2022 for NLDAS-2 and IT-Drought, and available record periods specific to each baseline forecast system shown in Extended Data Table 1. Following previous definitions of flash drought[14,17], we identified the start, onset phase, and termination of flash droughts based on the following criteria. A flash drought starts when the weekly SM decreases below the 40$^{th}$ percentile, followed by declining SM to the 20$^{th}$ percentile or lower with an average decline rate exceeding 5%. The onset phase of the flash drought terminates once the SM starts to increase or the average decline rate drops below 5%. The flash drought terminates when the SM percentile rises above the 20$^{th}$ percentile. And the duration of a flash drought event (from start to termination) must last 1 to 3 months.

**Forecast skill assessment**

We quantified the forecast skills of IT-Drought and the baseline forecast systems for SM and flash droughts, respectively, in each grid cell across CONUS. The forecast skill for SM was evaluated using the anomaly correlation coefficient (ACC) as a function of the forecast lead time $l$:

$$ACC_l = \frac{\sum_d \sum_y (Y_{y,d} - \bar{Y}_d)(P_{y,d,l} - \bar{P}_{d,l})}{\sqrt{\sum_d \sum_y (Y_{y,d} - \bar{Y}_d)^2 \sum_d \sum_y (P_{y,d,l} - \bar{P}_{d,l})^2}} \in [-1, 1] \qquad (2)$$

Here, $Y$ and $P$ denote the observed and the predicted SM values, respectively; $Y_{y,d}$ and $\bar{Y}_d$ are the SM in year $y$ and calendar day $d$ and its climatology, respectively, from NLDAS-2 product; $P_{y,d,l}$ and $\bar{P}_{d,l}$ are the forecasted SM in year $y$ day $d$ at the forecast lead time $l$ (from 1 to 46 days), and its climatology, respectively; $y$ denotes all years within the accuracy evaluation period; *d* denotes all days within the evaluation period when the forecast is available, i.e., 1–365 for IT-Drought, and days spaced by weekly forecast frequencies for the baseline forecast systems, depending on their forecast frequencies (Extended Data Table 1). The forecast skill for flash droughts was evaluated using the F1 score, i.e., the harmonic mean of precision and recall. We identified flash drought occurrence and absence based on the sequence of forecasted SM at a given lead time, following the above-described drought identification procedure. The binary occurrence was then used to calculate the F1 score by comparing it to drought occurrence detected from NLDAS-2. Both the ACC and the F1 score were calculated for IT-Drought and the baseline forecast systems. Given the different record periods across the baseline forecast systems, the forecast skills were quantified using the available periods close to the testing periods of IT-Drought for consistent comparison (Extended Data Table 1).

**Data availability**

The NLDAS-2 dataset is from [https://disc.gsfc.nasa.gov/datasets/NLDAS_NOAH0125_H_2.0/summary](https://disc.gsfc.nasa.gov/datasets/NLDAS_NOAH0125_H_2.0/summary); the MSWEP dataset is from [https://www.gloh2o.org/mswep/](https://www.gloh2o.org/mswep/); the ERA5-land dataset is from [https://cds.climate.copernicus.eu/datasets/reanalysis-era5-land?tab=overview](https://cds.climate.copernicus.eu/datasets/reanalysis-era5-land?tab=overview); the National Land Cover Database is from [https://www.usgs.gov/media/images/national-land-cover-database-nlcd-2021-conterminous-us-land-cover](https://www.usgs.gov/media/images/national-land-cover-database-nlcd-2021-conterminous-us-land-cover); the Shuttle Radar Topography Mission dataset is from [https://www.usgs.gov/centers/eros/science/usgs-eros-archive-digital-elevation-shuttle-radar-topography-mission-srtm-1](https://www.usgs.gov/centers/eros/science/usgs-eros-archive-digital-elevation-shuttle-radar-topography-mission-srtm-1); the Global Forest Canopy Height 2019 dataset is from [https://glad.umd.edu/dataset/gedi](https://glad.umd.edu/dataset/gedi);the GEFS reforecast dataset is from [https://www.emc.ncep.noaa.gov/emc/pages/numerical_forecast_systems/gefs.php](https://www.emc.ncep.noaa.gov/emc/pages/numerical_forecast_systems/gefs.php); the CMA, CNRM, ECMWF, and HMCR reforecast dataset is from [https://apps.ecmwf.int/datasets/data/s2s/.](https://apps.ecmwf.int/datasets/data/s2s/.)

**Code availability**

The source code for the IT-Drought model and reproducing figures and analyses in this paper are available at [https://github.com/qianzhao0408/flash_drought_forecast_interpret_paper](https://github.com/qianzhao0408/flash_drought_forecast_interpret_paper). The input data of the IT-Drought and reforecast datasets are public, and we don't provide them due to a lack of a data license. To run our model, you must obtain and pre-process the input data yourself. The data used to reproduce the figures reported in this paper are available at [https://doi.org/10.5281/zenodo.16903004](https://doi.org/10.5281/zenodo.16903004). The outputs of IT-Drought are available from the corresponding author on request. The source code is based on Python and MATLAB.

**Acknowledgments**

Y.L. acknowledges support from NASA grants 80NSSC22K1249 and 80NSSC21K1523, and from the Translational Data Analytics Institute at The Ohio State University.

**Author contributions**

Q.Z. and Y.L. designed the study. Q.Z. and X.T. developed the model. Q.Z. performed the research. Q.Z., X.T., X.Z., P.G., and Y.L. analyzed the results. Q.Z. and Y.L. wrote the initial draft. All authors edited the manuscript.

**Extended Data**

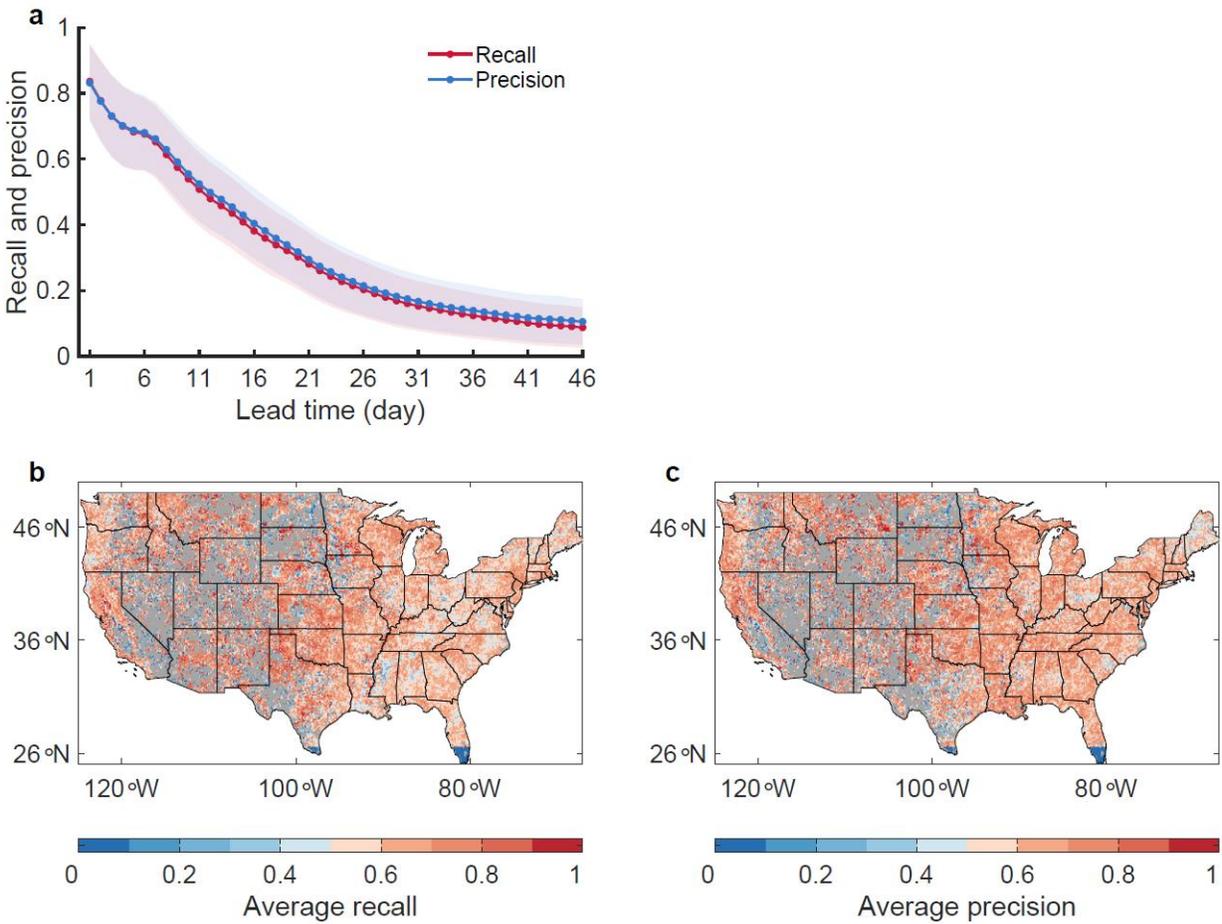

**Extended Data Fig. 1 | IT-Drought balances the recall and precision of forecasted flash drought occurrence**. **a,** The recall and precision of forecasted flash drought occurrence using IT-Drought at lead times from 1 to 46 days, quantified for the testing period (1997-2000 and 2019-2022). The spatial patterns of recall (**b**) and precision (**c**) of the flash drought forecast of averaged lead times of 1–14 days.

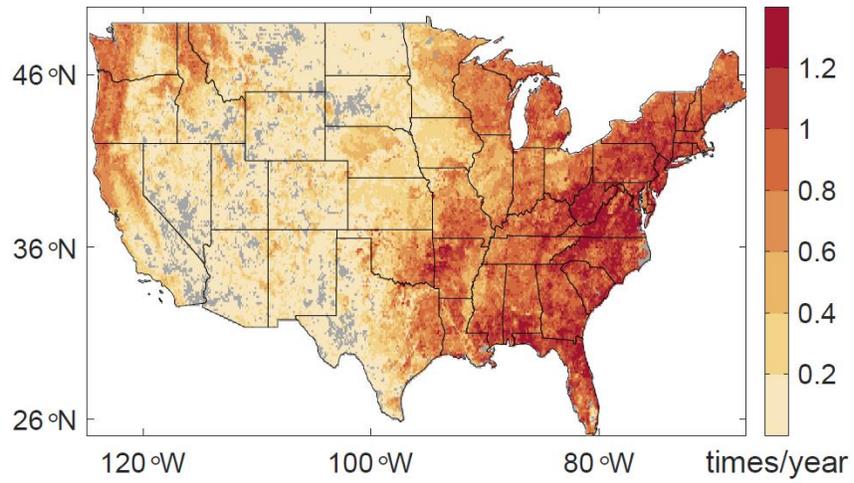

**Extended Data Fig. 2 | Frequency of flash droughts across the CONUS during 1979-2022.** Flash droughts are identified based on root-zone soil moisture percentile obtained from the NLDAS reanalysis datasets (Methods).

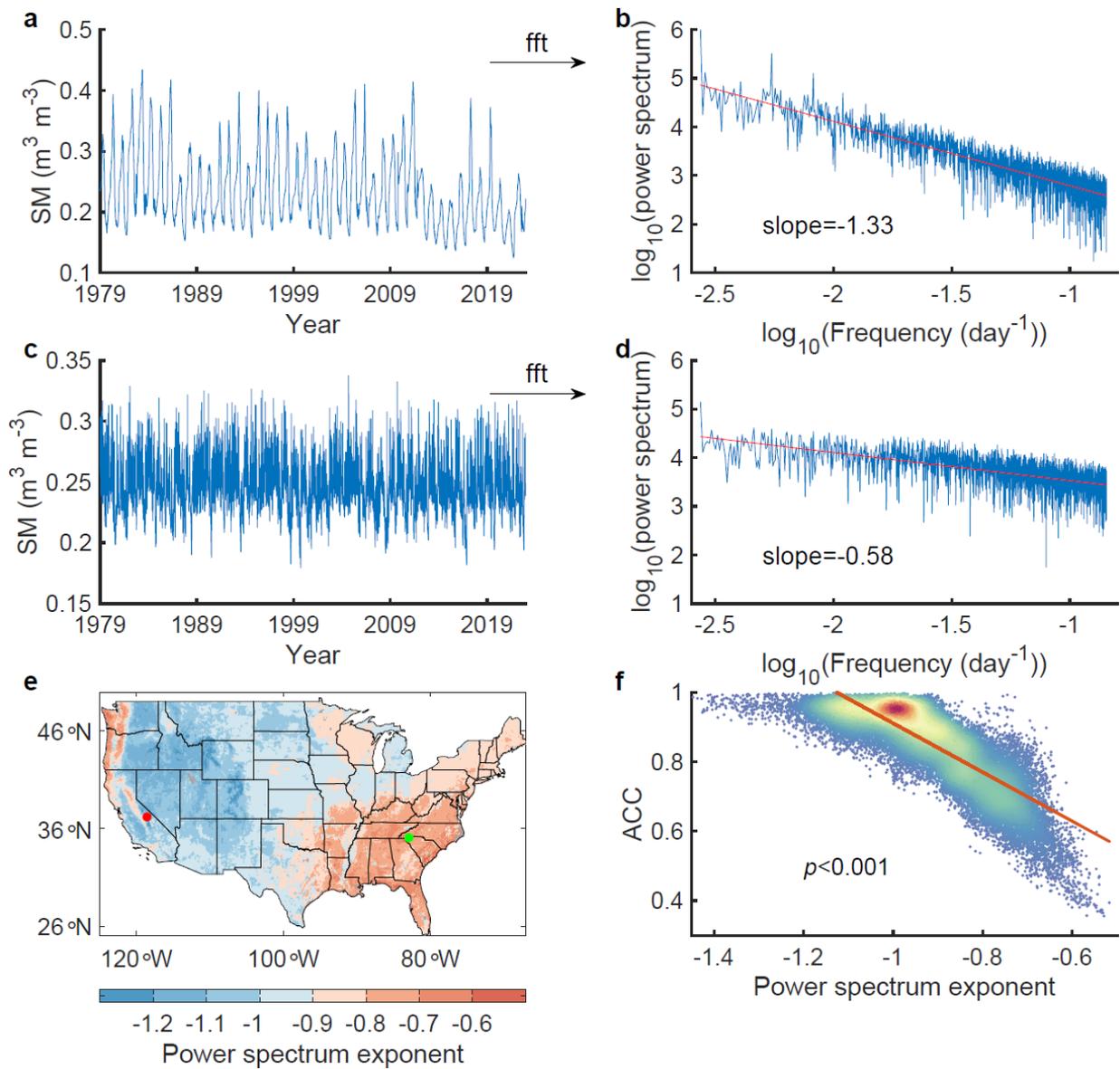

**Extended Data Fig. 3 | Soil moisture (SM) forecast accuracy tends to be lower in areas with faster-varying SM.** SM time series (**a, c**) and power spectra (**b, d**) contrasting fast-varying (**a, b**) and slow-varying (**c, d**) SM in two example grid cells in the Western US (red dot in **e**) and the Southeastern US (green dot in **e**). The power spectra are derived using the Fast Fourier Transform (fft), where the slopes of the regression lines (red lines) denote the exponents of the power spectra. **e,** The spatial pattern of the SM power spectrum exponent across CONUS. **f,** The relationship between the ACC of forecasted SM (Fig. 2c) and the power spectrum exponent (**f**)

across CONUS, with warmer colors representing higher dot densities. A negative regression slope (red line) suggests lower forecast accuracy in faster-varying regions.

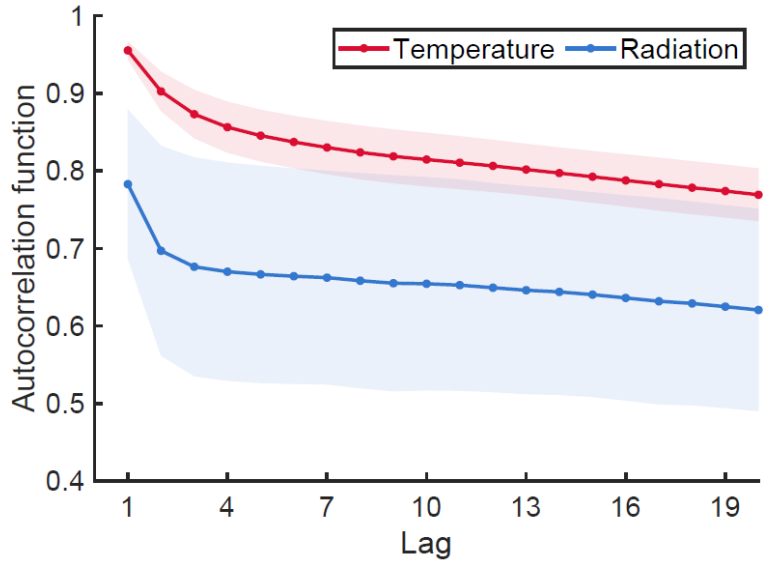

**Extended Data Fig. 4 | Temperature is more temporally autocorrelated than radiation.** The autocorrelations across time lags are calculated using the ERA5-land dataset product. The dotted lines and shaded areas denote the means and standard deviations across CONUS.

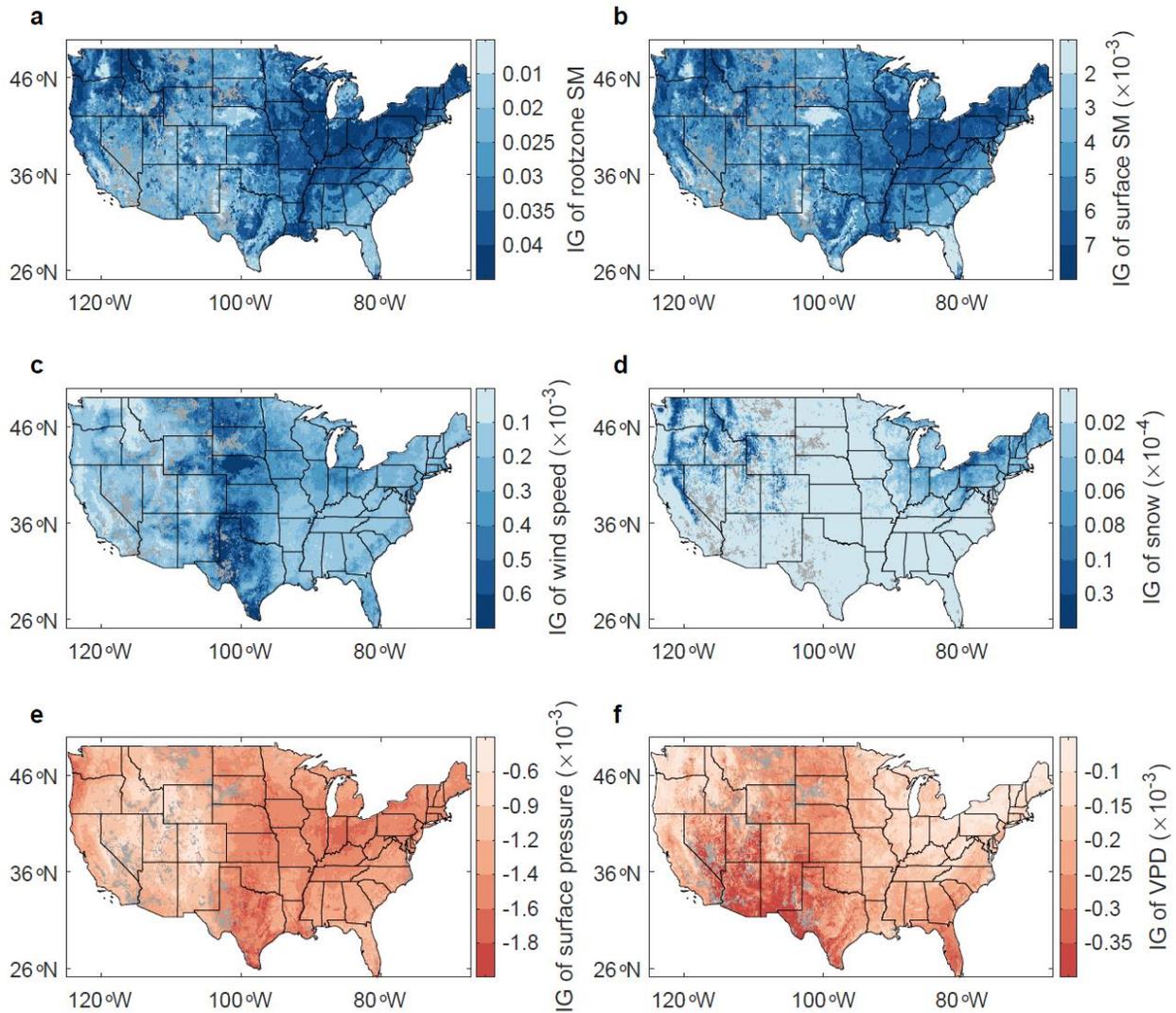

**Extended Data Fig. 5 | Root-zone soil moisture (SM) during flash drought onsets is positively affected by root-zone and surface SM, wind speed, and negatively by surface pressure and VPD.** The spatial patterns of Integrated Gradients (IGs) of root-zone SM (**a**), surface SM (**b**), wind speed (**c**), snow (**d**), surface pressure (**e**), and VPD (**f**) averaged over 7 days.

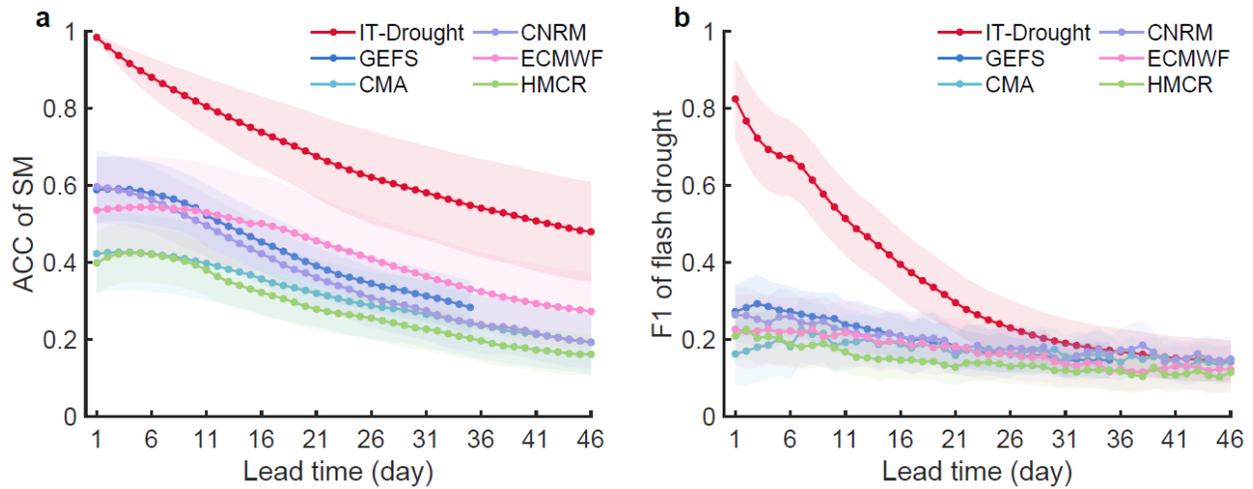

**Extended Data Fig. 6 | Evaluation of forecast systems against ERA5-land soil moisture (SM).** The anomaly correlation coefficient (ACC) of root-zone SM (**a**) and the F1 score of flash drought (**b**) forecasted by IT-Drought and five dynamic forecast systems (GEFS, ECMWF, CMA, HMCR, and CNRM) at lead times from 1 to 46 days. The ACC and F1 scores of forecast systems are calculated against the ERA5-land SM. The shaded areas show half of the standard deviations of the accuracies across CONUS. Both ACC and F1 are evaluated for the testing period (Extended Data Table 1).

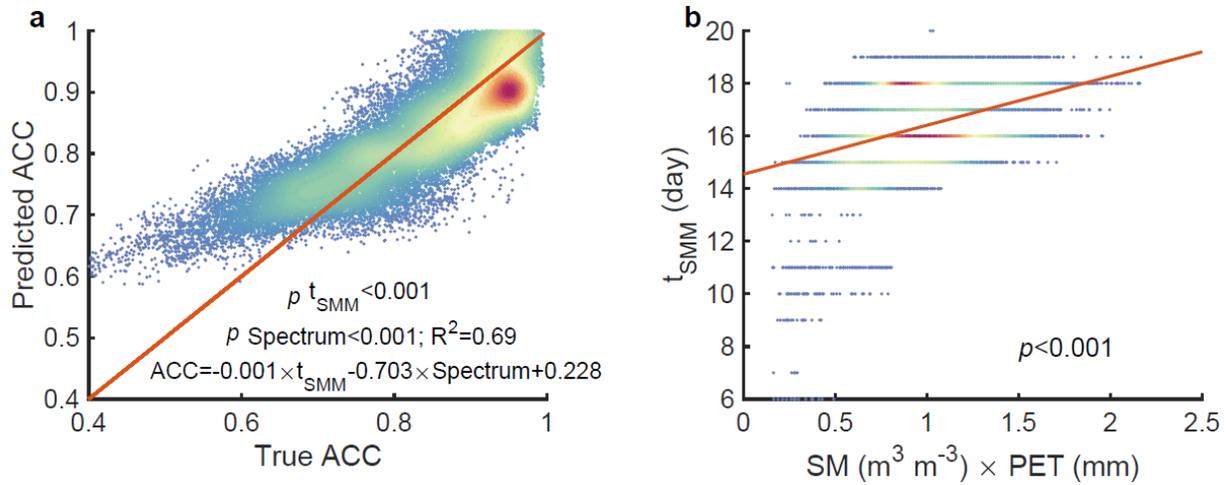

**Extended Data Fig. 7 | Across CONUS, soil moisture (SM) is forecasted more accurately in regions with lower spectrum exponents and shorter SM memory ($t_{SMM}$). a**. Short SM memory and low spectrum exponent of SM contribute to high ACC based on a multilinear regression model (red line). Each dot represents a grid cell with warmer colors denoting higher density. **b**. The relationship between the length of SM memory ($t_{SMM}$) and SM loss through evapotranspiration, approximated by the product of SM and potential evapotranspiration (PET) during drought onset. The red line indicates the fitted linear regression model ($R^2 = 0.13$, $p < 0.001$).

**Extended Data Table 1 | Dynamic forecast systems used as baselines for assessing forecast accuracies of soil moisture and flash droughts.**

| Model | Full name | Model version and date | Lead time (day) | Forecast updating frequency | Record period | Resolution | Period for accuracy evaluation |
|---|---|---|---|---|---|---|---|
| GEFS | Global Ensemble Forecast System | GEFSv12 | 0-35 | Weekly (Wed) | 2000-2019 | ~100 km | 2000-2003 and 2016-2019 |
| ECMWF | European Centre for Medium-Range Weather Forecasts | CY47R1 | 0-46 | 2/week (Mon/Thu) | 2000-2019 | ~100 km | 2000-2003 and 2016-2019 |
| CMA | China Meteorological Administration | BCC-CPS-S2Sv2 | 0-60 | 2/week (Mon/Thu) | 2005-2019 | ~100 km | 2005-2008 and 2016-2019 |
| CNRM | Centre National de Recherches Meteorologiques - Climate Model Version 6 | CNRM-CM 6.1 | 0-47 | weekly | 1993-2017 | ~100 km | 1997-2000 and 2014-2017 |
| HMCR | Hydro-Meteorological Centre of Russia | RUMS (2023) | 0-46 | Weekly (Thu) | 1991-2015 | ~100 km | 1997-2000 and 2012-2015 |
| IT-Drought | Deep-learning network with interpretability | - | 0-46 | Daily | 1979-2022 | 0.125° | 1997-2000 and 2019-2022 |